\documentclass[compsoc, conference, a4paper, 10pt]{IEEEtran}

\usepackage{paralist}

\usepackage{amsmath}
\usepackage[pdftex]{graphicx}
\usepackage{color}
\definecolor{light-gray}{gray}{0.75}

\usepackage{tabularx}
\usepackage{booktabs}
\usepackage{colortbl}
\usepackage[table]{xcolor}

\usepackage{listings}

\usepackage{hyperref}
\hypersetup{breaklinks=true}
\usepackage{subcaption}

\usepackage{eufrak}

\usepackage[normalem]{ulem}

\usepackage{footnote}
\usepackage[bottom]{footmisc}
\usepackage[utf8]{inputenc}
\usepackage{rotating}

\usepackage[T1]{fontenc} 
\usepackage[activate={true,nocompatibility},final,tracking=true,kerning=true,spacing=true,factor=1100,stretch=10,shrink=10]{microtype} 

\usepackage{xcolor}
\usepackage{xargs}

\usepackage{cite}
\usepackage{fixltx2e}

\usepackage{tikz}
\usetikzlibrary{calc, fit, positioning, decorations, decorations.pathreplacing}

\usepackage{amssymb}
\usepackage{pifont}
\newcommand{\cmark}{\ding{218}}%
\newcommand{\xmark}{\ding{217}}%
\newcommand{\dmark}{\ding{216}}%
\newcolumntype{L}{%
	>{\hspace{2ex}} X}
\newcolumntype{M}{%
	>{\bfseries\hspace{-2ex}
		\vphantom{\raisebox{1ex}{\rule{1ex}{1ex}}}}
	X
}
\newcolumntype{N}{%
	>{\bfseries\hspace{-1ex}
		\vphantom{\raisebox{1ex}{\rule{1ex}{1ex}}}}
	X
}

\DeclareMathOperator{\pk}{pk}
\DeclareMathOperator{\sk}{sk}

\DeclareMathOperator{\cycleid}{c_{\text{cid}}}

\newcommand{\cert}{\mathrm{Cert}}
\newcommand{\voucher}{\mathrm{Vch}}
\newcommand{\agency}{\mathrm{IN}}
\newcommand{\alice}{\mathrm{A}}

\begin{document}

\title{New Directions for Trust in the Certificate Authority Ecosystem}
\author{\IEEEauthorblockN{Jan-Ole Malchow, Benjamin Güldenring, Volker Roth\\ \tt\footnotesize <firstname>.<lastname>@fu-berlin.de}
	\IEEEauthorblockA{Freie Universität Berlin}}

\maketitle

\thispagestyle{empty}

\begin{abstract}
Many of the benefits we derive from the Internet require trust in the authenticity of HTTPS connections. Unfortunately, the public key certification ecosystem that underwrites this trust has failed us on numerous occasions. Towards an exploration of the root causes we present an update to the common knowledge about the Certificate Authority (CA) ecosystem. Based on our findings the certificate ecosystem currently undergoes a drastic transformation. Big steps towards ubiquitous encryption were made, however, on the expense of trust for authentication of communication partners. Furthermore we describe systemic problems rooted in misaligned incentives between players in the ecosystem. We depict that proposed security extensions do not correctly realign these incentives. As such we argue that it is worth considering alternative methods of authentication. As a first step in this direction we propose an insurance-based mechanism and we demonstrate that it is technically feasible.
\end{abstract}

\section{Introduction}
The Internet connects billions of individuals and countless organizations who use it every day for purposes ranging from mundane activities to transactions that warrant utmost confidentiality and integrity. Many of the benefits that Internet users enjoy hinge on trusting HTTPS connections and thus the Certificate Authority (CA) ecosystem.

We argue that the CA ecosystem has served us well over the years but is close to its end of life now. Obviously certificates are deeply integrated into technology in the field and will not vanish shortly. However, the trust to use them for authentication purposes is decreasing. We back this claim with two arguments. Our first argument is that the ecosystem underwent drastic changes in the last four years. These changes basically transformed the CA market into a gratis market in conjunction with a strong degree of centralization. On the one hand these changes increased the number of encrypted connections but on the other hand reduced the trust we can place on certificates. Our second argument is that even if the market still exists the inherent incentives are misaligned from a trust perspective. As a result we argue that it is worth considering alternative means of CA ecosystem variants.

Towards our first argument we present numbers from Google's Certificate Authority Server and an updated analysis of Durumeric et al. ``Analysis of the HTTPS Certificate Ecosystem'' from 2013~\cite{durumeric2013analysis}. Towards our second argument we follow and extend Arnbak et al.~\cite{Arnbak:2014:SCH:2668152.2673311} who already stated that CAs are ``too big to fail.''  We concur based on our own findings, but the underlying mechanics deserve more scrutiny. Existing efforts focused on informing decision making but overlooked the role of incentives. In this paper, we discuss the CA ecosystem from the perspective of incentives, striving to determine the effectiveness of checks and balances within that ecosystem. Based thereof, we argue that existing efforts meant to improve the CA ecosystem leave incentives largely as they are.  The implied conclusion is that existing proposals to improve the CA ecosystem might not be as effective in practice as we hope them to be as seen from a purely technical viewpoint.

Based on this foundation, we designed an alternative model that we call \emph{Connection Insurances (CI)}, and a proof of concept implementation.  A key difference between this model and prior work is that it alters the roles of the involved parties and the allocation of risks and rewards among them. Thereby, we manipulate the incentives of the actors in the ecosystem in order to adjust the system parameters.  In a nutshell, insurers assume the role of certification authorities.  Insurers are bound contractually to pay out benefits to Internet users in case they provide misleading information. We demonstrate that our design is technically feasible. On the flip side, we assume that Internet users pay for their security in the form of insurance premiums. Many readers will contend that this assumption is invalid in practice. However, our proposal is still a first step towards an alternative whose justification is not purely a technical one. We aim at demonstrating that a solution is possible by orchestrating well understood building blocks.

If we limit ourselves to purely technical considerations then we may end up with no sustainable solution at all.  This, however, is an undesirable outcome. Evaluation and analysis of our proof of concept implementation shows that our design is technically feasible. Our proof of concept requires only straightforward applications of well-understood cryptographic techniques, and is compatible to existing systems.

The rest of this paper is organized as follows. In \S~\ref{sec:ca_current} we present an updated analysis of the CA ecosystem. In \S~\ref{sec:technology_incentives} we discuss general misalignments in the CA system. In \S~\ref{sec:details} we describe the design with which we aim to mitigate these deficits, using insurances as a basic ingredient. In \S~\ref{section:evaluation} we evaluate the technical feasibility of our design. \S~\ref{sec:conclusions} concludes our work.

\section{Current State of the CA Ecosystem}
\label{sec:ca_current}
We compare the current numbers from Certificate Transparency~\cite{tack} (CT) and Censys~\cite{censys15} with the numbers reported by Durumeric et al.\cite{durumeric2013analysis} in 2013. For CT we obtained a log of Google's server~\cite{trans} Pilot on 2017-06-30 and removed all certificates that were expired or not yet valid.  We further removed 27 certificates with invalid UTF-8 entries in their \emph{Common Name} field.  This resulted in a dataset containing $31,487,506$ certificates. For Censys we used the most recent values as of 2017-06-30. We only took into account currently valid certificates. The resulting dataset contained $61,589,750$ certificates ($+682\,\%$ towards 2013).

Tables \ref{tab:ca_countries_top10} and \ref{tab:ca_top10} summarize the current state of the CA ecosystem regarding countries and CAs. Towards the countries issuing certificates the top countries have not changed. However, the United States further extended their leading role and now issues over $80\,\%$ of all certificates. The entire market further consolidated and over $95\,\%$ of all certificates are now issued by two to three countries, while in 2013 ten countries shared $95\,\%$. Moreover, China is now the country issuing fourth most certificates, while it was not in the top ten in 2013.

\begin{table}[t]
	\fontsize{9pt}{9pt}
	\centering
	\caption{\textbf{Top 10 Countries Issuing Certificates} -- All values are in percent. The issuer column corresponds to the respective certificate field.}
	\label{tab:ca_countries_top10}
	\begin{tabularx}{\columnwidth}{@{}X r r r r @{}}
		\toprule
		& CT & 	\multicolumn{2}{c}{Censys.io} & Durumeric~\cite{durumeric2013analysis}\\
		& Issuer & Issuer & EV Issuer & Issuer \\
		\midrule
		Austria         & 0.00   & 0.00   &  0.00 & 0.11   \\
		Korea	        & 0.01   & 0.01   &  0.00 & 0.24   \\
        Switzerland     & 0.09   & 0.12   &  0.16 & -      \\
        France          & 0.12   & 0.09   &  0.09 & 0.38   \\
        Poland          & 0.13   & 0.09   &  0.29 & -      \\
        Japan           & 0.14   & 0.15   &  2.06 & 1.06   \\
        Italy           & 0.14   & 0.09   &  0.11 & -      \\
        Germany         & 0.18   & 0.17   &  0.57 & 0.88   \\
        Netherlands     & 0.22   & 0.23   &  3.67 & 1.32   \\
        China           & 0.37   & 0.66   &  0.19 & -      \\
        Israel          & 0.38   & 0.86   &  0.93 & 2.56   \\
        Belgium         & 1.55   & 1.89   &  4.39 & 3.29   \\
        United Kingdom  & 16.35  & 10.8   & 15.37 & 10.88  \\
        United States   & 80.11  & 81.83  & 69.03 & 77.55  \\
        \midrule
        Sum             & 99.58  & 96.99  & 96.86 & 98.27  \\
		\bottomrule
	\end{tabularx}
\end{table}

While the distribution among countries is only slightly changed (see Tab.~\ref{tab:ca_countries_top10}) the issuing CAs changed drastically (see Tab.~\ref{tab:ca_top10}). The foremost change is that \emph{Let's Encrypt} has a market share of more than $50\,\%$, while it not even existed in 2013. In addition the former market leader \emph{Symantec Corporation} was outperformed by multiple other CAs especially by its direct competitors \emph{COMODO Group} and \emph{GoDaddy.com Inc.}. At the time of writing it is expected that Symantec will sell its certificate business to \emph{DigiCert Inc.}, this will however not change the relation of market shares. Another major change is that \emph{cPanel Inc}, \emph{Western Digital Corporation} and \emph{Amazon Inc.} are now among the top ten issuers.

\begin{table*}[htb]
	\fontsize{9pt}{9pt}
	\centering
	\caption{\textbf{Top Certificates and Parent Companies} -- This table combines the top 10 values for all three data sources. The percent values are summed up for the respective parent company.}
	\label{tab:ca_top10}
	\begin{tabularx}{\textwidth}{@{}X l r r r r@{}}
		\toprule
		\textbf{Certificate Authority} & \textbf{Parent Company} & Certificate Transparency & Censys.io & Censys.io EV & Durumeric et al.\cite{durumeric2013analysis}\\
		\midrule
        Amazon & Amazon, Inc.                                       & 0.48\,\%   & 0.38\,\%  & -          & -\\
        COMODO CA Limited & COMODO Group                            & 18.21\,\%  & 11.12\,\% & 15.37 \,\% & 11.80\,\%\\
        cPanel, Inc. & cPanel, Inc.                                 & 11.83\,\%  & 7.51\,\%  & -          & -\\
        Entrust, Inc. & Datacard Group                              & 0.24\,\%   & 0.24\,\%  &            & 2.40\,\%\\
        GeoTrust Inc. & DigiCert, Inc.                              &            &           &            & -\\
        thawte, Inc. & DigiCert, Inc.                               &            &           &            & \\
		Symantec Corporation & DigiCert, Inc.                       &            &           &            & 34.23\,\%\\
        DigiCert Inc. & DigiCert, Inc.                              & 6,46\,\%   & 9.79\,\%  & 54.77 \,\% & 4.19\,\%\\
        TERENA & GEANT Association                                  & 0.18\,\%   & 0.14\,\%  &  3.02 \,\% & 1.22\,\% \\
        GlobalSign nv-sa &  GMO Internet, Inc.                      & 1.74\,\%   & 1.95\,\%  &  4.39 \,\% & 4.90\,\%\\
        GoDaddy.com, Inc. & GoDaddy.com, Inc.                       & 3.89\,\%   & 2.96\,\%  &  7.48 \,\% & 29.13\,\%\\
        Network Solutions &  Web.com Group, Inc.                    & 0.14\,\%   & -         &  -         & 1.81\,\% \\
        Let's Encrypt & Internet Security Research Group            & 53.48\,\%  & 61.21\,\% &  -         & - \\
        Western Digital Tech. & Western Digital Corporation         & -          & 1.59\,\%  &  -         & - \\
        StartCom Ltd. & WoSign CA Limited                           & 0.71\,\%   & 1.22\,\%  &  0.93 \,\% & 2.56\,\%\\
        \midrule
        Sum                &                                        & 97.21\,\%  & 99.09\,\% & 85.96 \,\% & 98.65\,\%\\

		\bottomrule&
	\end{tabularx}
\end{table*}

\subsection{Extended Validation Certificates}
Extended validation (EV) certificates require a more rigorous vetting than other certificates. A separate, comparably small market with $637,911$ certificates in the field (Censys, 2017-09-27) emerged. The countries issuing domain validated (DV) certificates are also the major players in the extended validation market. The top four issuing countries accounting for $88.79\,\%$ are identical. Six major companies in the DV market also account for $85.96\,\%$ of the issued EV certificates, with DigiCert issuing more than $50\,\%$. The market is slightly more diverse than the DV market, however, both are dominated by a single player.

\section{Technology versus Incentives}
\label{sec:technology_incentives}

In this section we review systemic aspects of the existing public key certification system and several technical research proposals meant to augment or replace it.

\subsection{General Misalignments}
\label{subsec:miss_alignment}
In a well-balanced system, all parties have a choice, and money and trust flow in the same direction.  The reason simply is that when trust is betrayed, the flow of money stops. This serves as a disincentive that counters the desire to betray trust in order to maximize some other influx of money. For this basic economic mechanism to function three criteria must be met. First there must be a money flow that can be altered. Second the money flow must be aligned to trust such that it can be used as a leverage. The choice to deny further payment must not be without alternative. If for example one can not use the entire Internet as a consequence of money withdraw one practically does not have leverage.

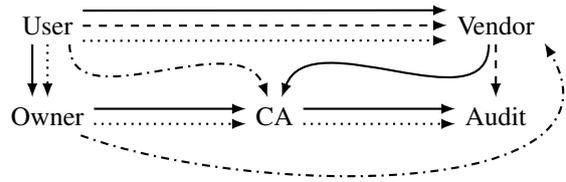
\begin{figure}[t]
	\centering
	\begin{tikzpicture}
	[
	node distance=0.7cm and 2cm,
	>=latex, thick, font=\fontsize{9pt}{9pt}
	]
	\node (Benutzer) {User};
	\node[below= of Benutzer] (Domain)     {Owner};
	\node[right= of Domain]   (CA)         {CA};
	\node[right= of CA]       (Audit)      {Audit};
	\node[above= of Audit]    (Hersteller) {Vendor};

	\draw[->, dashed] ([yshift=0.0 cm] Benutzer.east)     -- ([yshift= 0.0cm]Hersteller.west);
	\draw[->]         ([yshift=0.2 cm]Benutzer.east)      -- ([yshift=0.2 cm]Hersteller.west);
	\draw[->, dotted] ([yshift=-0.2 cm]Benutzer.east)     -- ([yshift= -0.2 cm]Hersteller.west);
	\draw[->]         ([xshift=-0.2 cm]Benutzer.south)    -- ([xshift=-0.2cm]Domain.north);
	\draw[->, dotted] (Benutzer.south)                    -- (Domain.north);
	\draw[->, dashdotted] ([xshift=0.2 cm]Benutzer.290) to[out=-90, in=110] (CA);

	\draw[->]                   ([yshift=0.1 cm]Domain.east)    -- ([yshift=0.1 cm]CA.west);
	\draw[->, dotted]           ([yshift=-0.1 cm]Domain.east)   -- ([yshift=-0.1 cm]CA.west);
	\draw[->, dashdotted](Domain.330) to[out=-20, in=300] (Hersteller.340);

	\draw[->]         ([yshift=0.1 cm]CA.east)  -- ([yshift=0.1 cm]Audit.west);
	\draw[->, dotted] ([yshift=-0.1 cm]CA.east) -- ([yshift=-0.1 cm]Audit.west);

	\draw[->, dashed] (Hersteller) -- (Audit);
	\draw[->]         (Hersteller.250) to[out=-90, in=70] coordinate (m) (CA);
	\end{tikzpicture}
	\caption{\textbf{Misaligned Incentives} -- Choice (solid), money (dotted), trust / dependency (dashed) and implicit trust / dependency (dashdotted) in the CA model (including extended validation).}
	\label{fig:trust_money}
\end{figure}

Consider Fig.~\ref{fig:trust_money}, which depicts the flow of money, choice and trust among the players in the CA ecosystem, namely the users who surf the web, the vendors of browsers, the domain owners, the certificate authorities and their auditors. Fig.~\ref{fig:trust_money} depicts that implicit trust relations (dashdotted lines) exist that are not aligned with choice or money flow. As a consequence users do not have a leverage against misbehaving CAs. More than 200 CAs worldwide are operative and cam potentially issue rogue certificates. Towards the browser vendors choice and money of users are aligned. However, users can, depending on their operating system, choose from two out of three CA lists (Mozilla, Microsoft and Apple). As a result the user is practically very limited in his choice. He thus does not possess a practical leverage against browser vendors. Domain owners posses a leverage against the CA that issued their certificate. This is why CAs offer so called ``Protection Plans'' for their certificates, that basically insure that the CA does not misbehave against its customers. However, there exists an implicit trust relation with browser vendors (dashdotted line). This relation is not aligned with choice or money and as such the domain owner does not have a leverage here.

\subsection{Analysis of Systemic Factors}
\label{subsec:sym_fac}
Klitgaard published a formula to describe the basic mechanics of opposing incentives in~\cite{klitgaard1988controlling}. His pseudo formula equates \emph{corruption} $(\mathfrak{C})$ with \emph{monopoly} $(M)$, \emph{discretion} $(D)$ and \emph{accountability} $(A)$ as follows: $\mathfrak{C} = M + D - A$. Klitgaard initially introduced this formula to describe the abuse of power, namely corruption. However, in the context of this paper the formula has to be understood as a generalized description of tension between incentives. As we do not imply that any participant is in fact corruptible we will use the term \emph{tension} in the following.

Informally, one may affect the level of tension in a system by reducing monopoly or discretion, or by increasing accountability. The three parameters allow us to describe which parameters a proposed approach alters. Furthermore, the formula describes the relation between the parameters. This allows us to evaluate if the values for the parameters chosen by a proposal are well aligned. If they are not, the proposal does not fully solve the problem.

At the time of writing, researchers primarily used technical parameters as an appraisal of the effectiveness of their proposed solutions, for example, the time to recover from a certificate breach. With the application of Klitgaard's equation the effectiveness of a proposed solution can be answered from a systemic viewpoint. Currently overlooked limitations or weaknesses in proposed solutions can be determined. In what follows, we look at the CA system by means of these three parameters.  Since we do not have direct indicators of discretion and accountability for the CA ecosystem, we must infer them from what we know about the relationships of the market participants as described in \S~\ref{subsec:miss_alignment}. Please keep in mind that the following description refers to the basic model of CAs, extensions like Certificate Transparency are discussed later.

\subsubsection*{Monopoly}
As shown in \S~\ref{sec:ca_current} the CA ecosystem (DV and EV) is centralized with one major participant (see Tab.~\ref{tab:ca_top10}). With the numbers reported by Durumeric et al.~\cite{durumeric2013analysis} in 2013 as starting point this development showcases the formation of a monopoly. A major factor is that DV certificates are a digital commodity with practically zero marginal costs. In combination with the requirement for ubiquitous encryption the development towards a centralized gratis market is logical for DV certificates. In fact there is no mechanism in place to prevent the forming of monopolies in the DV or EV CA ecosystem.

\subsubsection*{Discretion}
Discretion means that a business entity does not have to coordinate with or seek approval of another independent entity when making business decisions. CAs generally do not need to seek approval of anyone prior to issue a certificate. Later sanctions for issuing certificates is covered under accountability.

\subsubsection*{Accountability}
\label{subsec:accountability}
An entity is accountable if another independent one oversees it and if it faces a consequence for misbehavior that is suitable to deter it from misbehaving repeatedly. Based on an analysis of the version history of Mozilla's NSS security module we found that, Mozilla delisted or distrusted 13 certificates between 2008 and the mid 2017, however more incidents occurred.  On four occasions, the security of a CA was breached by hackers.  On 193 occasions, a CA issued certificates unauthorized. On 1614 occasions certificates for not registered domains where issued. On four occasions, an issued certificate was in violation of a baseline requirement or policy. None CAs delisted or distrusted in the aftermath of an incident was among the top 10 market shareholders. Rather, Mozilla seeks to delist or distrust certificates as far down the certification hierarchy as possible in order to minimize the impact on users~\cite{mozilla_dcssi,caBan2016}. Symantec and its subsidiaries issued hundreds of rogue certificates (including EV certificates) and were not distrusted but forced to use Certificate Transparency~\cite{symantec_issues}. Only after another year of not complying to policies Google and Mozilla laid out a multi-phase plan to remove Symantec~\cite{symantec_revoke}. The bottom line is that holding top tier CAs accountable is challenging for other players in the CA ecosystem. In addition the CA market leader is operated by a consortium including the major vendors. As such they are essentially their own watchmen now.

\subsection{Application to Existing Work}
\label{subsec:related}
\label{subsubsec:incentives}

Approaches of existing work can be described as different methods to adjust the parameters monopoly, discretion and accountability. In the following we group proposals by the addressed parameter.

\subsubsection*{Reduction of Monopoly}
\label{subsec:distribution}

Distribution of control is used to reduce monopolies and thus lowers tension. The underlying assumption is that not all controlling entities are corruptible or that subverting multiple entities is too difficult or costly. Proposals that rely on the distribution of control are AKI~\cite{kim2013accountable}, ARPKI~\cite{Basin:2014:AAR:2660267.2660298}, Certificate Cothority~\cite{syta2015certificate} and PoliCert~\cite{Szalachowski:2014:PSF:2660267.2660355}.  However, the mechanisms of choice and trustworthiness remain similar to those in place for the established CA system.  Domain owners select by market penetration and price, browser vendors select by market penetration and industry self-regulation criteria, and users select by convenience, that is, they continue to follow defaults and vendors' suggestions. Trust on First Use (TOFU) and TACK~\cite{tack} remove third parties completely and as such monopolies are not a concern. HSTS Preloading~\cite{rfc6797} is managed by the browser vendor who becomes a monopolist, taking into consideration the limited choice consumers have.

\subsubsection*{Reduction of Discretion}
\label{subsec:discretion}
DANE ~\cite{rfc6698} and Cage~\cite{kasten2013cage} aim at reducing discretion of CAs. The underlying idea is to restrict CAs so that they can issue certificates for selected domains only. The necessary policies are either derived from the existing distribution of certificates (Cage) or defined by the domain owners, as in DANE. \emph{DANE} introduces an additional root of trust. From this root, trust propagates along the delegations of domains to sponsoring organizations. VeriSign is the assigned sponsor in the case of \texttt{.com} domains~\cite{iana_root}.  Since VeriSign is part of the Symantec Group (expected to be owned by DigiCert by 2018), a major player in the CA market, the consequential lack of independence of DANE and the CA market limits the benefits derived from DANE. Trust on First Use (TOFU) and TACK~\cite{tack} remove third parties completely as such discretion is not existing. HSTS Preloading~\cite{rfc6797} is managed by the browser vendor and thus increases discretion. TOFU and TACK would be ideal from the perspective of tension resistance but they provide limited protection in the cases of first-time website visits, loss of pinning state and, likely, habituation effects due to repeated acceptance of first-time keys.

\subsubsection*{Increase of Accountability}
\label{subsec:blaming}
Perspectives~\cite{wendlandt2008perspectives}, Convergence~\cite{conv}, Sovereign Keys~\cite{sover}, Certificate Transparency~\cite{trans}, AKI~\cite{kim2013accountable}, ARPKI~\cite{Basin:2014:AAR:2660267.2660298}, PKI Safe Net~\cite{Szalachowski:2016aa} and PoliCert~\cite{Szalachowski:2014:PSF:2660267.2660355} all use public logging in one form or another to make misbehavior of CAs visible. \emph{PKI Safe Net} was designed especially to address the \emph{too-big-to-fail} problem of major CAs~\cite{Szalachowski:2016aa}. The design allows to revoke CAs at specified points in time in order to not break the entire system. The assumption is that this eases revocation decisions for browser vendors.  All mentioned proposals assume that the threat of disclosure of misbehavior is a sufficient deterrent for repeated misbehavior. However, there is evidence that this is a rather strong assumption in the case of CAs~\cite{Arnbak:2014:SCH:2668152.2673311}, as we elaborated before in Section~\ref{subsec:accountability}.

\section{Connection Insurances for Certification}
\label{sec:details}
We introduce the concept of \emph{Connection Insurances (CI)}. The following description of the system follows a top down approach first we define and specialize desired properties and then describe instantiations on the next detail level. The approach is meant to demonstrate the orchestration of well understood techniques to achieve the desired properties, rather than being a final solution. We especially do not address problems arising from flaws in the TLS protocol or current implementations in browsers.

The core intent of connection insurance (CI) is that Internet users, who face the most direct consequences of rogue certificates, decide for themselves whom to trust and they benefit from holding their trust anchors accountable. For sake of simplicity we leave  domain owners out, they can, however, be integrated as proxies for insurance policies. In order to hold someone accountable some prerequisites are required: All partners must know each other, all partners must voluntary agree to the arrangement, there must be some stakes at risk in case of misbehavior and all partners must be in the same jurisdiction.

The typical approach is to construct an arrangement that is enforceable by law, a contract. Using CAs to authenticate communication partners is a risk mitigation strategy. If we assume that the risk can not be nullified there remains a chance of damage to the user. The classical approach to cope with inevitable risks is delegation and distribution of the risks, namely an insurance. Based on this we explored the design space for an insurance backed system.

An insurance is generally free in its decision what and who to ensure. As such discretion can not be altered by our approach. According to the described mechanics (\S~\ref{subsec:sym_fac}) we need to keep monopoly low and increase accountability. The basic idea of insurances is that they are accountable if an insured event occurs. As such accountability is a basic property of such an approach. If the user can select from a variety of insurance companies there is no monopoly. Based on this idea we seek to construct a system that allows insurances for authenticity of connections.

In the proposed model users select their anchor of trust, that is their insurer. As such all partners know each other and voluntary agree. As users select their issuer they can select one in the same jurisdiction. As a result the user can rely on the jurisdiction if a conflict with the insurer appears. We proceed with a high level description of how CI work and perform with respect to Klitgaard's formula.

\subsection{Design of Relations}
From a security protocol perspective, CI involves the following parties: A user \emph{Alice} who visits web sites and who is insured by an \emph{insurer} $\agency$. $\agency$ maintains a list of trustworthy certificates. $\agency$ gets to know new certificates by means of application or own search. In either case a certificate will most likely undergo a harsh vetting process. This is, as for any insurance, part of the risk management process of $\agency$. When Alice visits a web server \emph{Bob,} information provided by her insurer informs her whether the certificate $\cert_{\text{Bob}}$ presented by Bob is trustworthy or not. We additionally assume a legal trusted third party \emph{Judge} $J$ who is responsible for settling disputes between Alice and $\agency$ (if one of them misbehaves) and who enforces legal contracts signed by Alice and $\agency$.
We will say an \emph{insurance case} occurs if all of the following events occur at the same time $t$:
Alice connects to Bob's web server, Bob's server presents a certificate $\cert_{\text{Bob}}$, $\agency$ told Alice that $\cert_{\text{Bob}}$ is trustworthy and $\cert_{\text{Bob}}$ is a rogue certificate.
We say that a certificate is \emph{rogue} if the identity in the certificate is not Bob's.  These cases may occur, for example, if someone: forges a certificate in the name of a trusted CA or steals a web server's private key
and subsequently uses the corresponding private key in a MITM attack on Alice.  Our mechanisms prevent case~1 and indemnify Alice in case~2. We do not require that Alice knows at the time she connects to Bob that $\cert_{\text{Bob}}$ is rogue but rather that she learns this at some later point in time $t' > t$ (for example by E.U. notification requirements as mentioned in~\cite{Szalachowski:2016aa}). If an insurance case occurs, we say that Alice is eligible for compensation or, equivalently, that $\agency$ is liable.  If $\agency$ is liable then, by contractual arrangement, $\agency$ pays a benefit to Alice. The amounts of premiums and benefits are determined by the market.

\subsubsection*{Tension Mechanics}
In the following we evaluate for each party in the system whether they have an incentive to misbehave. We relay on the introduced mechanics of tension for our evaluation.

\noindent
\emph{Alice} -- Alice, as single user, is not in a position of power regarding the overall market. As such Alice is not corruptible in terms of our definition.

\noindent
\emph{Bob} -- Operates his domain.
\emph{Monopoly}:  \dmark ~ Bob can not decide about the $\agency$ Alice choses. As such there exists no monopoly nor has Bob discretionary power.
\emph{Discretion}: \xmark ~ This parameter remains untouched.
\emph{Accountability}: \cmark ~ If Bob misbehaves, Alice and $\agency$ can hold him accountable. This is especially to not have any further business relations.
In terms of the tension formula this is an adequate situation where, without monopoly, accountability nullifies existing discretion.

\noindent
\emph{Insurer} -- $\agency$ collects rewards and is the trusted third party for authentication.
\emph{Monopoly}: \xmark ~ Customers are free to choose their trusted
	parties independently. There is no system immanent mechanism that limits their choice.
\emph{Discretion}: \xmark ~ This parameter remains untouched.
\emph{Accountability}: \cmark ~ The design of our protocols assures that
	customers can demonstrate their insurance claims rigorously to a judge.
There is no system immanent mechanism that supports monopolies. Discretion is not changed as described in (\S~\ref{sec:details}). However, accountability is increased what nullifies the existing discretion. Based on introduced mechanics this constellation prevents tension.

\noindent
\emph{Vendor} -- Vendors including the current CA list into their product are not involved in the decision making process in CI.

\noindent
\emph{Judge} -- We assume that judge, in the form of judiciary, has an intrinsic motivation to be honest and fair.

\subsection{Threat Model}
Our system is designed to protect Alice and her insurance provider $\agency$
from misbehavior of each other. In particular, it provides CI to Alice:
\emph{If the insurer promised Alice that certificate $C$ is trustworthy at time $t$ and Alice established a connection to a web server who presented $C$ at time $t$ then Alice can provide convincing evidence of this fact to a judge.}
Since we require that Alice's evidence is convincing this also protects
$\agency$ from fraud (that is, Alice cannot falsely claim that an insurance
case occurred).

Additionally, our system provides \emph{reselling protection} (RP) to the
insurer.  Without RP, Alice may sell her policy file to a third party, say,
Charlie.  This does not extend Alice's coverage to Charlie but it provides
Charlie with protection against certificate substitutions.  This results in
a free-riding problem for the insurance market.  While we cannot prevent
Alice from sharing information she receives from the insurer we can assure
that this information is not convincing to a third party:
\emph{Alice cannot convince others that her insurer considers a certificate to be trustworthy without explicit interaction of her insurer.}
At first sight, CI and RP appear contradictory. However, the contradiction
can be resolved by using \emph{Chameleon signatures}~\cite{krawczyk1998chameleon2} in lieu of
conventional signatures.

In an interactive protocol between Alice and her insurer there may be
situations where one party benefits from deviating from the protocol.  In
general, we allow Alice to attempt fraud and to abort protocol instances.
We require that the insurer does not abort the protocol, though  (for example
by denying one of the signatures in \emph{Certificate List Download} below).
The
insurer may however refuse payment of benefits and may prepare such a
refusal by sending crafted messages.  In summary, we accept that Alice may
be malicious, but require the insurer runs the protocol to completion.

\subsection{Insurance System}

In what follows, we succinctly explain the protocols we designed to support
CI. The goal of our construction is to demonstrate that CI can be implemented using well-understood cryptographic techniques. We begin with a simplified version of CI, which lacks privacy, efficiency
and reselling protection but is easy to understand.  We add said properties subsequently.

Alice initially receives a list of (trusted) certificates from her insurance
provider and keeps the list up to date by checking regularly for
updates. Whenever she connects to a web server she creates a \emph{voucher}
that the server signs. Before Alice updates her certificate list she submits
the accumulated vouchers (one per domain) to her insurance provider and
receives a receipt for the submission. If one of her trusted certificates is
compromised then she presents the corresponding voucher and receipt to her
insurer and claims the insurance benefit.  We now describe the details of
the scheme.

Our scheme uses a public key signature scheme $\Pi$, a collision resistant
hash function $h$ and a cryptographic hash function $H$ modeled as random
oracle.  We write $\sigma_A(m)$ to denote a signature of $A$ on $m$ where
$A$ may be a name (for example, Alice) or a certificate. If we write
$\sigma_A(m, m')$ then this implies that $m$ and $m'$ are padded to suitable
lengths and concatenated afterwards.

\noindent
\emph{Insurance Provider Setup} -- In order to set up service, the insurer creates a signature key pair
$(\pk_{\agency}, \sk_{\agency})$ and assembles an initial list of
certificates $C := (\cert_1, \dots)$ that it believes to be authentic.

\noindent
\emph{User registration} -- In order to subscribe to the insurer's service, Alice creates a signature
key pair $(\pk_A, \sk_A)$ and her insurer generates a unique customer number
for Alice. For simplicity we equate this number with Alice.  The insurance
contract between Alice and her insurer includes the public keys
$(\pk_{\agency}, \pk_{\alice})$ of both, the customer number of Alice, the
contract's validity term $(t_0, t_{\operatorname{end}})$ and an upper bound
$\Delta T$ on the time between certificate updates.  If Alice does not
update her certificates within the bound then she forfeits her coverage
until the next update occurs.  Both parties must agree to the contract in a
legally binding way.

\noindent
\emph{Certificate List Download} -- The certificate list download marks the beginning of an update cycle. The insurer sends a list of
certificates $C$ and a randomly generated cycle identifier $\cycleid$ to
Alice.  Alice computes a signature on $\cycleid, C$, a timestamp $t$ and her
customer number and sends the signature to the insurer.  The insurer
verifies Alice's signature, checks that the timestamp is recent and replies
with a signature of the same message.

As we will see later, the insurer has to make sure it issues a unique $\cycleid$
to Alice.  Reusing a $\cycleid$ would enable Alice to claim compensation for
a previously issued list of certificates. If $\cycleid$ is randomly chosen and sufficiently
long, say 256 bit, the probability of reusing a previous $\cycleid$ is negligible.

\noindent
\emph{Voucher Submission} -- The voucher submission marks the end of an update cycle and occurs right
before the beginning of the next certificate list download.  Alice assembles
a list of vouchers $V$ and sends the hash value $h(V)$ and a timestamp
$t'$ to the insurer together with a signature of these values, the current
cycle identifier and her customer number. Her insurance provider checks the
signature, that the timestamp is recent and that Alice did not violate the
update interval requirement.  The insurer then replies with a signature of
the same message.
Our simplified scheme requires that Alice and the insurer store all messages
and signatures exchanged in each cycle.

\noindent
\emph{Creating Vouchers} -- A voucher provides evidence of the fact that Alice established a connection
to Bob's server at \texttt{bob.example.org} whose certificate
$\cert_{\text{Bob}}$ was included in certificate list $C$ in update cycle
$\cycleid$. Alice computes vouchers as follows: she generates a fresh random
value $r$, sets
$v := \langle \text{Alice}, \texttt{bob.example.org}, \cycleid, r \rangle$
and calculates $H(v)$. She sends $H(v)$ to Bob who replies with a signature
$\sigma_{\text{Bob}}(H(v))$. Alice stores the tuple
$\voucher := \langle \cert_{\text{Bob}} , v, \sigma_{\text{Bob}}(H(v))\rangle$
where $\voucher$ is her voucher.

Ordinary web servers are not aware of this
protocol. However, during an TLS ephemeral key exchange servers sign a
client-chosen \texttt{random\_bytes} value, where we can put a randomized
version of $H(v)$.

\noindent
\emph{Demonstrating Insurance Cases} --
\label{sec:easy_dispute}
Assume that Bob's certificate $\cert_{\text{Bob}}$ turns out to be rogue,
that Alice connected to Bob's server, that Alice received
$\cert_{\text{Bob}}$ and that Alice wishes to claim a benefit for this case.
In order to support her claim, Alice has to prove three things:
1) $\cert_{\text{Bob}}$ was in the certificate list $C$ in cycle $\cycleid$,
2) Alice updated her certificate list in time, and
3) Alice connected to Bob's server in cycle $\cycleid$.
Alice proves the first two items by disclosing certificate list $C$, cycle
id $\cycleid$ timestamps $t,t'$ and the signatures
$\sigma_{\agency}(\text{Alice}, \text{``Certificates''}, \cycleid, t, C)$
and
$\sigma_{\agency}(\text{Alice}, \text{``Vouchers''}, \cycleid, t', h(V))$
issued by $\agency$ in cycle $\cycleid$.  Verifying that
$\cert_{\text{Bob}}\in C$ is straightforward.  Updates were timely if $t'-t
\leq \Delta T$.  The evidence is convincing because $\agency$ chose a unique
cycle id $\cycleid$ and Alice cannot compute the signatures herself without
breaking the signature scheme. Neither can Alice reuse signatures issued to
another customer because signatures of the insurer include unique customer
numbers.

In order to prove the third item, Alice discloses the voucher list $V$ that
matches hash value $h(V)$ and the corresponding voucher $\voucher$.
Verifying that $\voucher\in V$ is straightforward.  The proof is convincing
because Alice would have to forge a valid signature of Bob or find a
collision in one of the hash functions $h$ or $H$ in order to compute the
evidence herself.  Since $v$ includes her customer number she cannot reuse a
voucher from another customer.

\subsection{Optimizations}
In the simplified scheme, Alice discloses her entire list of vouchers in order to claim compensation.  This leaks connection information to the insurer.  Also, our description above does not yet offer reselling protection as described before and is not very efficient in terms of storage, in that both Alice and her insurer have to remember the details of sent and received signatures.

\subsubsection*{Voucher Privacy}
In the simplified scheme, Alice discloses her entire list of vouchers in order to claim compensation.  This leaks connection information to the insurer.  We can solve this problem by making $h(V)$ the root of a Merkle tree.  In order to prove that a voucher is a leaf $\ell$ of the tree, Alice discloses the path from the root to $\ell$.

This still leaks the order of $\ell$ in $V$ and the size of $V$.  We can avoid this leakage by making the size of $V$ equal to the size of $C$.  Note that $|C|$ is an upper bound of $|V|$.  Alice simply pads $V$ to the size of $C$ using pseudorandomly generated vouchers and randomizes the order in $V$.  She does not have to store the entire padded tree because she can recreate it at any time from a randomly chosen seed.

\subsubsection*{Reselling Protection} \label{sec:reselling}
The simplified scheme allows Alice to resell her certificate list to others. All she needs to do is disclose $C, t, \cycleid$, her customer number and the signature of the insurer.  If Alice does not want to disclose the signature than she can still prove in zero-knowledge that she knows a valid signature.  This is sufficient to convince third parties that the insurer considers the certificates in $C$ valid and authentic.  In order to remove this property, we require that Alice and the insurer use a \emph{Chameleon signature scheme}~\cite{krawczyk1998chameleon2} in lieu of a regular signature scheme.  Chameleon signature schemes have the property that Alice can convince herself that the insurer signed a message but she cannot convince others that the insurer is the signer because Alice could have forged the signature herself. The construction in~\cite{krawczyk1998chameleon2} makes use of a Chameleon hash function $\mathcal{H}(\cdot,\cdot)$ in combination with a regular signature scheme.

In order to uncover forgeries, the insurer is now required to record all
messages signed in this fashion.  If the insurer uses an authenticated
channel to send this message or the certificate list to Alice then the
authentication must be deniable.  Otherwise, Alice can use a transcript of
the communication in lieu of regular signatures in order to convince third
parties about the authenticity of the certificate list.  A technicality of
the Chameleon construction is that Alice must prove knowledge of a trapdoor information ($\tau$) when
signing the contract with the insurer~\cite{krawczyk1998chameleon2} (see
also~\cite{jakobsson1996designated}).  Otherwise, Alice may employ a notary
to create the Chameleon hash function for her (without disclosing $\tau$).

\subsubsection*{Storage Efficiency}
We can reduce Alice's storage requirements as follows. Instead of storing
$C$ for every cycle, Alice only stores the up-to-date certificate list $C$
and keeps a log of changes to preceding updates.  More precisely, let $C_i$
be the certificate list after the $i$-th update cycle. The update log
consists of a list of roll-back functions $\Delta(\cdot,\cdot)$ that, given
$C_i$, output $\Delta(C_i, i) = C_{i-1}$. If some $C_j$ expired (because the
policy term does not cover it any longer) then the corresponding roll-back
functions can be deleted.

\subsubsection*{Integration with TLS}
We focus on the ephemeral key exchange methods \texttt{DHE\_DSS} and \texttt{DHE\_RSA} in RFC~5246 because \texttt{DH\_anon} does not authenticate the server and the non-ephemeral key exchange methods do not provide forward secrecy. The TLS~1.2 handshake begins with a \texttt{ClientHello} message from the client to the server. This message contains a field called \texttt{random\_bytes} with 28 bytes that the client chooses. The server replies with a \texttt{ServerKeyExchange} message containing a signed structure called \texttt{signed\_params}. This structure contains the random values from \texttt{random\_bytes} that the client chose before. When Alice visits Bob's website she sets \texttt{random\_bytes} to $H(v)$. After receiving this message from Bob, she extracts the signature \texttt{signed\_params} and stores the result.

\subsection{Related Models}
Other researchers have suggested before that the certification of keys
should be tied to insurances.  Reiter et al.~\cite{reiter1999authentication}
made this the basis of an authentication metric. The authors draw analogies to the reinsurance business but their work remains theoretical and focused on metrics.

More recently, Matsumoto and Reischuk~\cite{matsumoto2015certificates} made
a case for tying certificates with insurances as a means to improve the
accountability of CAs.  The model they propose differs from ours in crucial
aspects.  Matsumoto and Reischuk suggest that customers (browsers) hold CAs
accountable for the issuance of rogue certificates.  As a consequence, CAs
are required to pay an insurance benefit to the domain owners whose domain
was the subject of the rogue certificate. While we aim to insure the end user.

\section{Technical Feasibility}
\label{section:evaluation}
The technical feasibility depends on three parameters, connection overhead, performance overhead and storage requirements. Where the last said is the most important as usage of local storage is the main technical difference compared to existing approaches. The actual TLS connection remains unchanged as such there is no direct connection overhead. There is an additional connection for the first occurrence of a certificate in an update cycle, this however, amortizes over time. In terms of computation overhead one signature creation and one signature validation per update cycle are required. As such these overhead amortizes over time as well. We implemented Chameleon signatures in the Go programming language and ran benchmarks on a MacBook Pro 2012 (Mac OS X 10.11.3, 2.7\,GHz Intel Core i7, 16\,GB). It took $\approx 0.4\,\text{ms}$ to compute on average and verifying them took $\approx 0.5\,\text{ms}$ on average. As such even the delay for the initial connection is below human perceptibility and hence we predict that an introduction would not be noticeable to users.

\subsubsection*{Storage requirements}
In what follows, we estimate the amount of storage that insurers and their customers need in order to run the CI system. Both need to store the generated vouchers and users need to store encountered certificates. In the absence of better approaches we rely on existing statistics of parameters that we cannot easily collect ourselves. We define that informations must be stored for 365 days this definition is based on the fact that the average time to fix a 0-day exploit is 321 days \cite{Bilge:2012:BWK:2382196.2382284}.

All encountered certificates $n$, with average size $s_{cert}$ must be stored for $d$ times. Where $d$ equates as $d = 365 / k$ where $k$ is the average validity time of certificates. Thus the required storage capacity $c_{cert}$ equates as $c = n \cdot s_{cert} \cdot (365 / k - 1) \approx 2.7\,\text{GB}/\text{year}$.
{$n=5\cdot 10^5$ websites generate $97\,\%$ of global e-commerce revenue~\cite{rjmetrics}. We use this as an upper bound to the number of different domains a user visits.
{$s_{cert}=1.90\,\text{kB}$:} Based on the dataset described in \S~\ref{sec:ca_current} $\approx 95\,\%$ of all certificates were smaller than 1.90\,kilobytes.
{$k=90$:} Based on the dataset described in \S~\ref{sec:ca_current} we found that $\approx 53\,\%$ of all certificates are valid 90 days.

All created vouchers $v$ of size $s_{voucher}$ need to be stored for one year.
{$v=2,500$:} By analyzing 1 million web sites in January 2016, Englehardt~\cite{Englehardt2016} recently observed that only 123 third party domains are present on more than $1\,\%$ of sites. We first selected the Alexa top 200 and found in summary less than 1000 domains.  The absolute number increased when selecting 200 sites at random, but still giving on average 1000 (ranging from $920$ to $1070$ in 20 repetitions). When counting subdomains separately, the number increased to 2426 for the set of  Alexa top 200 web pages, and to 1912 on average for randomly selected sites (ranging from $1578$ to $2103$ in 20 repetitions). In summary, between 1000 and 2500 certificates seem to be a reasonably estimate of the number of certificates Internet users encounter per day on average. We have no means to estimate how long an insurer would want a cycle to be but~1 day seems reasonable. Therefore $v = 2,500$ for the customer.
{$s_{voucher}$:} The required structures require $1,700$ bytes per voucher for the customer and $512$ bytes for the insurer. There are $128$ additional bytes per update cycle.
{$c_{voucher}$:} For the customer $2,500 \cdot 1,700 + 128 \cdot 24 \approx 0.58\,\text{GB} / \text{year}$. We assume that insurers have about 44 million customers (the total number of car insurances of the largest car insurer in the U.S.). This leads to a storage requirement of $512 \cdot 24 \cdot 365 \cdot (44 \cdot 10^6) \approx 180\,\text{TB} / \text{year}$.

Despite our conservative estimates (60 minutes update cycles, $5\cdot 10^5$ websites, $2,500$ Certificates), the storage overhead for users is reasonable, given the amounts of storage typically available on contemporary devices. We overestimated the required storage of insurers as well. Still, our conservative estimate appears reasonable, given current storage prices. Therefore, we conclude that CI is feasible from a storage perspective for both insurers and their customers.

\section{Conclusion}
\label{sec:conclusions}
The success of Let's Encrypt transformed the certificate market into a gratis market. Therefore the structure of competitors changed since the last review of the market in 2013. We argue that relying the security of the entire World Wide Web on just a hand full of companies rooted in a single country is potentially critical. Currently corporations are serving us well. However, there is no system inherent reason for this. Corporations are solely driven by the need to make money. Currently making money is aligned with a good user experience, that includes secure connections. But this alignment is not part of the system design itself, there is no general penalty if security is not the primary focus. Moreover, recent political events like the last US presidential election or the Brexit in Europe (some $90\,\%+$ of certificates are issued in these two countries) made clear that the status quo is not immutable.

In a considerable body of work, researchers pointed out shortcomings and failures of the CA ecosystem and made suggestions how it may be improved at a technical level. By applying a notion for the alignment of incentives we teased out immanent problems of the system, though. We especially showed that existing work overlooked the incentives of participants in the system. We showed that the parameters \emph{monopoly}, \emph{discretion} and \emph{accountability} can be used to identify the system immanent incentives. Based thereof, we argue that technical improvements alone are insufficient to address the inherent problems of the CA ecosystem. Consequently, we explored new directions for establishing a trust ecosystem.

Towards this end, we proposed a model that we call \emph{Connection Insurances}. This model puts users into the center of the system and enables them to source public key information from insurers they choose and trust and who are willing to commit to the authenticity of the information they provide.  In cases of misbehavior, users are able to prove their claims in a court of law and are eligible to collect benefits. Our model decreases \emph{monopoly} while increasing \emph{accountability}. In our analysis and evaluation, we found that our proposal is feasible in terms of computational and storage requirements. Overall, we believe that our model offers ample opportunity for further investigation and we hope that our first steps inspire other researchers to come up with improvements.

\bibliographystyle{IEEEtranS}

\end{document}